\documentclass[12pt]{iopart}
\usepackage{iopams}
\usepackage[pdftex]{graphicx}
\begin{document}

\newcommand{\Esca}{\mathcal{E}}

\title{Motivating quantum field theory: the boosted particle in a box}

\author{Amar C. Vutha}
\address{York University, Toronto, Canada}
\ead{avutha@yorku.ca}
%\pacno{} 	%

\begin{abstract}
It is a maxim often stated, yet rarely illustrated, that the combination of special relativity and quantum mechanics necessarily leads to quantum field theory. An elementary illustration is provided, using the familiar particle in a box, boosted to relativistic speeds. It is shown that quantum fluctuations of momentum lead to energy fluctuations, that are inexplicable without a framework that endows the vacuum with dynamical degrees of freedom and allows particle creation/annihilation.
\end{abstract}

Quantum field theory (QFT) is a consistent quantum theory of particle creation and annihilation \cite{Weinberg2005, Zee2010}. In its relativistically covariant formulation, QFT is the basis of the Standard Model, our best model yet for the innermost workings of the universe. Introductory courses in QFT typically begin with classical field theory, leading on to the Lagrangian description and building up to scalar field theory and quantum electrodynamics. Somewhere along this development, the claim is usually made that QFT is the only consistent framework for combining quantum mechanics and special relativity. However, a simple example that exhibits the dissonance between quantum mechanics and relativity is seldom provided. It is the purpose of this article to provide such a simple example, using concepts that are familiar to physics undergraduates. The example highlights the fundamental role played by quantum fluctuations. It is hoped that it might provide some physical intuition for students, and a simple paradigm for instructors, to better motivate the ensuing heavy mathematical machinery of QFT.

\begin{figure}[btp]
\centering
\includegraphics[width=0.9\textwidth]{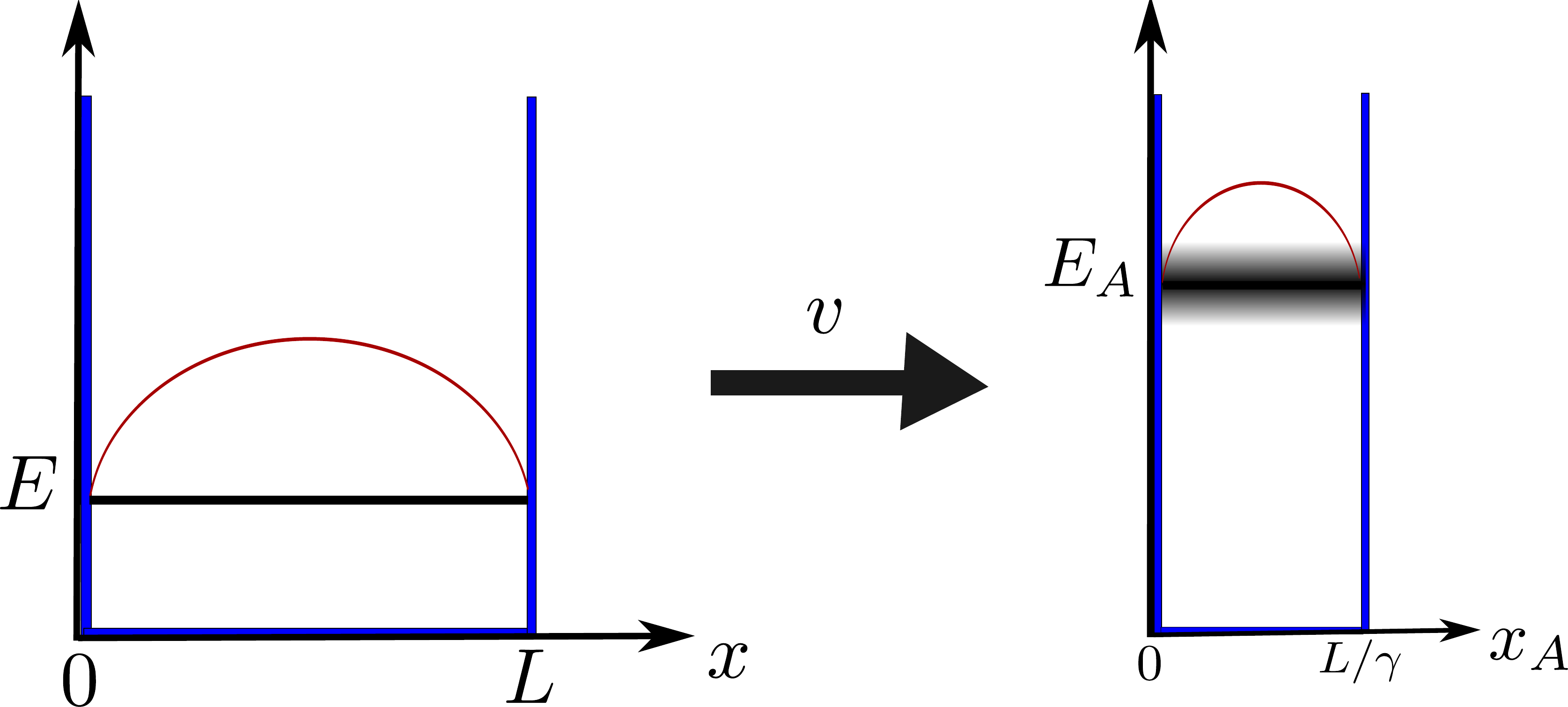}
\caption{A particle in a box of length $L$. The wavefunction of the ground state is schematically drawn (in red). According to an observer $A$ who finds it moving with a velocity $v$, the box appears to be Lorentz contracted and the energy of the ground state correspondingly raised. However, the energy of the ground state also exhibits quantum fluctuations, unlike the situation in the rest frame of the box.}
\label{fig:one}
\end{figure}

Consider the textbook situation (\emph{e.g.} \cite{Fitzpatrick2010}) of a particle of mass $m$, in a box (an infinitely high 1-dimensional potential barrier) of length $L$ as shown in Figure \ref{fig:one}. The ground state of this system has the wavefunction $\psi_0 = \sqrt{ \frac{2}{L}} \sin(\pi x/L)$. The momentum operator $\hat{p}$ has the expectation value $\langle \psi_0 | \hat{p} | \psi_0 \rangle = 0$, as expected of a particle that is at rest in the box. Equivalently, the $\sin(kx)$ term in the wavefunction can be written as the sum of traveling waves, $e^{i k x}$, with opposite momenta $k = \pm \hbar \pi/L$. The expectation value of the square of the momentum is $\langle \psi_0 | \hat{p}^2 | \psi_0 \rangle \equiv p_{qf}^2 = (\hbar \pi/L)^2$. The particle has a fluctuating momentum $\Delta p =  p_{qf}$, where these fluctuations of order $\sim \hbar/L$ are due to the confinement of the particle to a region of size $L$ -- in other words, these are just the quantum fluctuations of the momentum as a result of the uncertainty principle. 

We will explore the properties of the ground state as the box is boosted relativistically (see \emph{e.g.} \cite{Rindler1991} for an introduction to relativity). (In the following, it is assumed that lengths are measured in time-delay units, so that velocities are dimensionless and $c$ = 1.) Referred to the zero of energy at the bottom of the potential, the energy of the ground state is $E = \Big(\frac{\hbar\pi}{L}\Big)^2 /2m = p_{qf}^2/2m$, where $m$ is the rest mass. Combining this with the mass-energy of the particle, $m$, the correct relativistic expression for the energy of the ground state is $E =  \sqrt{ m^2 +  p_{qf}^2 }$. The energy-momentum four-vector of the particle in its ground state, in the box's frame, is therefore
\begin{equation}
(E_{box},p_{box}) =  ( \sqrt{m^2 + p_{qf}^2}, \pm  p_{qf} ).
\end{equation}
The momentum is written as $\pm p_{qf}$ to emphasize that this is a fluctuating quantity. Note that the energy-momentum has the expected dispersion relation: $E_{box}^2 - p_{box}^2 = m^2$.

Now consider the situation according to an observer $A$, in whose frame the box is moving at a velocity $v$. The energy-momentum of the particle in the box according to $A$ is
\begin{equation}
(E_A,p_A) =  \gamma \ \Big( \sqrt{ m^2 +  p_{qf}^2} \mp  v p_{qf},  \pm p_{qf} - v  \sqrt{ m^2 +  p_{qf}^2 }    \Big),
\end{equation}
where $\gamma = 1/\sqrt{1-v^2}$. The energy-momentum according to the observer $A$ once again has the correct dispersion relation, as can be verified after some simple algebra:  $E_{A}^2 - p_{A}^2 = m^2$.

The energy of the particle has the mean value $\langle E_A \rangle =\gamma \sqrt{ m^2 +  p_{qf}^2}$. This is what might be naively expected by $A$,  being equal (for small $p_{qf}/m$) to the apparent mass-energy of the relativistic particle, $\gamma m$, plus the ground state energy of a particle in a Lorentz-contracted box, $\frac{\hbar^2 \pi^2}{2 \gamma m} \frac{\gamma^2}{L^2}$. However, $E_A$ contains a fluctuating term as well. According to the observer $A$, the uncertainty in the energy is 
\begin{equation}
\Delta E_A = \sqrt{ \langle E_A^2 \rangle - \langle E_A \rangle^2 } = \gamma v p_{qf} \label{eq:deltaE}.
\end{equation}
From the point of view of relativity, there is no real mystery here: the fluctuations in the momentum have been converted, due to the boost, into fluctuations in the energy. This is just what would happen for a classical ensemble of particles too, when the momenta of the particles are randomly set to $\pm p_{qf}$. 

However, there is only a single particle in the box, and the fluctuations are quantum mechanical in origin. From the point of view of quantum mechanics, something quite bizarre has happened: the ground state, which is an energy eigenstate (with well-defined energy) in the box's frame, has a fluctuating energy according to $A$ and can no longer be properly called an energy eigenstate! It is not out of the ordinary for excited states to have a finite energy uncertainty, as a result of tunneling or decays to lower energy states. However, the situation according to $A$ is very odd, because the \emph{ground} state of this completely confined system has a finite energy uncertainty, and therefore a finite lifetime due to the energy-time uncertainty relation. Energy ought to be conserved, so where does this fluctuating energy go? What is the reservoir with which this energy is being exchanged? There is no real answer to this problem within the framework of the Schrodinger equation, as the ``vacuum'' is not a dynamical entity in this framework. This conundrum would be a fine place to stop and point out that the formulation of quantum mechanics in terms of the Schrodinger equation is not consistent with relativity, and move on to the building blocks of QFT. However, one further point can perhaps be made. 

Assume that this fluctuating energy is indeed being exchanged with the vacuum. What would the size of the momentum fluctuations $p_{qf}$ need to be, in order for this fluctuating energy to be greater than $m$ i.e. what are the conditions under which this fluctuating energy can potentially result in a new particle? Setting $\Delta E_A \geq  m$ in Equation \eref{eq:deltaE} implies that
\begin{equation}
p_{qf} = \frac{\hbar \pi}{L} \geq m \sqrt{ \frac{1}{v^2} - 1 }.
\end{equation}
As the box's velocity $v$ approaches 1, $A$ finds that arbitrarily small quantum fluctuations (arbitrarily large values for the confinement length $L$) result in energy fluctuations large enough to create a particle of mass $m$.

All of this highlights the need for a consistent quantum theory of particle creation/annihilation, which treats the vacuum as a dynamical entity. \emph{Ergo} QFT.

\section*{Acknowledgments}
I acknowledge spirited discussions with Kosuke Kato and Nikita Bezginov that led to the search for a simple example. This work was supported by York University.

\section*{References}
\bibliography{qft}
\end{document}